# Scalable Production and Supply Chain of Diamond using Microwave Plasma: a Mini-review

Sergey V. Baryshev[1,2] and Matthias Muehle[3,1]

[1]Department of Electrical and Computer Engineering, Michigan State University, MI 48824, USA

[2]Department of Chemical Engineering and Material Science, Michigan State University, MI 48824, USA

[3]Fraunhofer USA, Inc. Center Midwest, East Lansing, Michigan 48824, USA

**Abstract**

Discovered and reported exactly 40 years ago, microwave plasma assisted chemical vapor deposition (MPACVD) pointed out an economic technology that could potentially produce lab-grown diamond stones at scale. After this breakthrough discovery, demonstrating that diamond can be growth at low pressure and temperature, the progress quickly curbed and synthetic single crystal diamond (SCD) size and quality could not be improved toward attaining requirements critical in solid-state electronics. This led to the early promise of MPACVD to not come true and slowed the level of investments, thereby further stalling the progress in diamond syntheses. With the invention of a few novel homo- and hetero-epitaxy growth techniques, the diamond research and technology has recently reinvigorated. This mini review attempts to capture the momentum of recent progress in diamond MPACVD that could finally bring scalable manufacturing of high quality large size wafers for future electronics and optics.

**Introduction**

Like no other gemstone, diamonds have been praised in human history and culture. Diamonds were scarce, and thus a symbol of wealth and prestige. Beyond jewelry, it quickly became clear that diamond possesses extraordinary mechanical, thermal, and electronic properties and has been deemed the *ultimate engineering material and frontier material of the 21st century*. Indeed, diamond has the potential to reshape a wealth of applications ranging from quantum computing to electric grid infrastructure. Radiation hardness makes diamond the detector for space and high energy physics [1-3], and for high flux X-ray optics [4] and sensing [5, 6]. Diamond's resilience to high electric fields and superior thermal properties makes it attractive for generating electron beams in high pulsed power systems [7, 8]. Optically active vacancy center states in diamond make it a viable platform for quantum computing [9, 10]. An exceptionally large electrochemical window makes diamond an electrode of choice for photo-electro-chemical treatments of strongly bonded chemicals [11]. Diamond is biocompatible, thus enabling infinite opportunities in biology and medicine through implantable electronics [12, 13]. Several key figures of merit exist [14-18] showing that diamond outnumbers any existing semiconductor when used for power applications. It is illustrated in great detail in Fig.1.

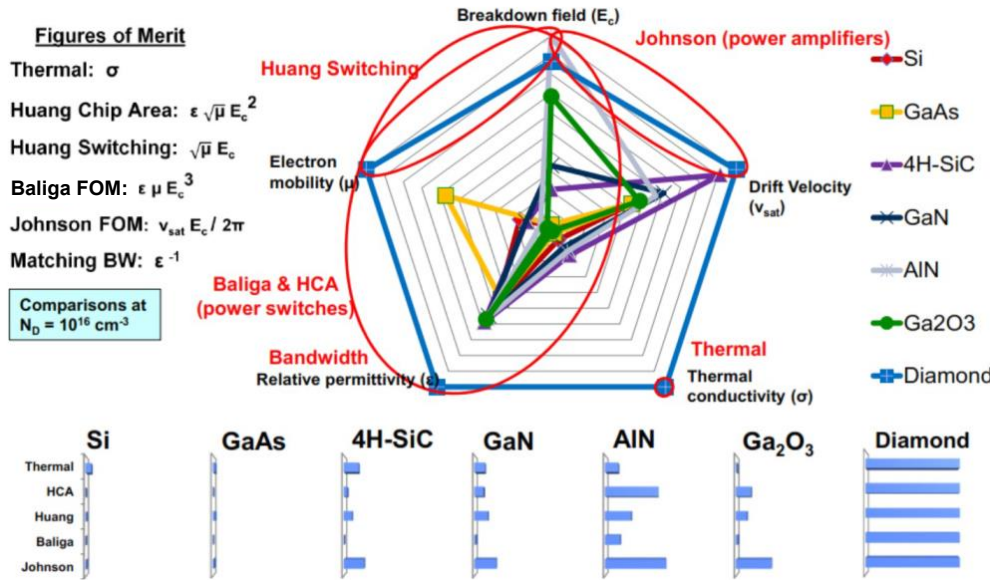

Fig.1. Key electronics figures of merit compared for major semiconductors. Adapted from Ref.[19].

To realize the promise that diamond holds for future electronics, high quality, large size diamond plates must be produced at scale. Developed for plethora of semiconductors [20], the main focus of the Czochralski's method was on the carbon group of the periodical table of elements. Discovered for tin, the method was quickly developed and perfected across the globe for Ge, leading to creation of the BJT at Bell Labs, and Si, fueling the third industrial revolution. Si wafers of perfect quality are available in up to 300 mm diameters. Because of the allotropy of carbon, the Czochralski's method could not be simply adapted to synthesize diamond. Historical efforts at General Electric Company [21] made it possible to grow diamond under high-pressure, high-temperature (HPHT) conditions similar to those found deep underground. Diamond seed would pull a new crystal of a few mm from a molten carbon feedstock kept at 10-15 GPa (equivalent to the pressure at ~200 km below ground level) and 2,000-3,000 K in a press chamber twice an average human height. Such conditions facilitate the reverse graphite-to-diamond transition (Fig.2), where the $sp^2$ hybridized graphite is the fundamental ground state of carbon and $sp^3$ hybridized diamond is a metastable state of carbon.

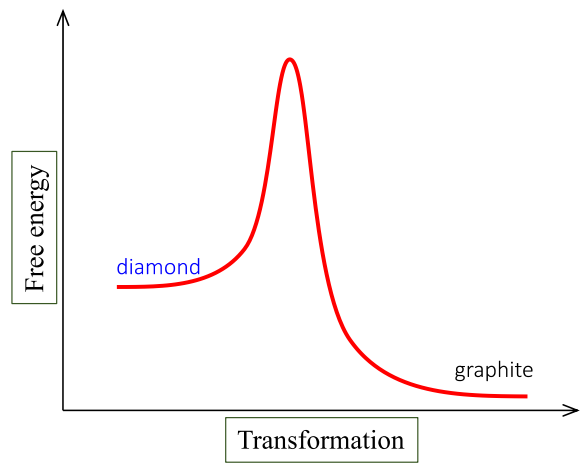

Fig.2. Elementary view of $sp^3$ to $sp^2$ transformation.

The material nearest to the seed is usually poor in quality, and the highest quality diamond plates are cut from the outer regions of the boule, meaning that the boules must be much larger than the desired plates. The growth rates of crystal-clear type IIa HPHT diamond with N-concentration of ~10 ppb must be very slow to prevent the uptake of nitrogen and other impurities. In Ref.[22], one large size diamond plate 7×5 mm$^2$ was extracted from a 10 carat HPHT boule that took (sic!) 11 days to grow. Producing large size diamond plates via HPHT is deemed unpractical, and at this point restricts the supply chain for future microelectronics. The long time standard high-quality diamond plate in HPHT technology available on the marketplaces has been 3.5×3.5 mm$^2$ in size.

Chemical vapor deposition (CVD) of diamond at subatmospheric pressures then came as a remedy. By different accounts, it is claimed that diamond CVD was pioneered in USA [23] or in the Soviet Union [24] in the 1960's. There is a patent by William G. Eversole of Union Carbide Corporation applied in 1958 [25] that precedes the work by Angus and Spitsyn. In this patent, three important discoveries were claimed: 1) methane was used as a major feedstock for synthesis, 2) hydrogen was successfully used for diamond product cleaning from co-deposited black carbon by-product, and 3) all syntheses were successful at 1,000 °C. These three major findings set the stage for what followed as diamond CVD breakthroughs. CVD methods differ and named according to a particular way reactive species, methyl and atomic hydrogen, are activated: hot filament [26], arc [27], combustion flame [28], glow discharge [29] and, finally, microwave plasma assisted CVD (MPACVD) [30] are mainly known due to their efficiency in growing diamond. MPACVD stood out over the past 40 decades as the only means capable of producing ultrapure single crystal diamond (SCD) at relative scale in terms of the lateral size and thickness and quantity. The other CVD methods have shortcomings in terms of product chemical impurity cleanliness, crystallinity or scalability.

To grow diamond, one needs diamond to initiate the growth. Historically, the choke points for MPACVD were always the seed crystals that had to be sourced from HPHT grown substrates, and therefore historically the SCD production by CVD revolved around crystals of no more than 3-5 mm is linear size. With no special means, in homoepitaxy, the film being grown is as large and good as the initial seed/substrate. The initial substrate must have the lowest possible defect densities, starting with point defects in terms of desirable impurity levels and line defects in terms of dislocation densities. As a rule, dislocations must be reduced as low as possible to make the best performing devices. Plane and bulk defects are just unacceptable. Simultaneously, useful substrate area must be as large as possible for industry to even consider diamond as a base material. The initial dislocation densities tend to be the lowest in the HPHT technology. Substrates with defect densities as low as $10^3$ cm$^{-2}$ are readily available (Fig.3). This value is at least an order of magnitude lower than any other type of single crystal diamond on the market, and several orders of magnitude lower than any natural diamond [31]. Because HPHT crystals are small, maintaining or further lowering impurity concentration was usually the main challenge addressed in MPACVD. Even so, optically clear, and so-called electronic grade CVD diamond plates appear black when viewed by X-rays because the number of dislocations is vast with a typical lateral density on the order of $10^5$ 1/cm$^2$. Fig.3 shows that CVD diamond quality is very inconsistent when compared between different groups that produce them (5 orders of magnitude spread), where dislocation density below $10^3$ 1/cm$^2$ is rarely achieved. Such high dislocation densities negate diamond's supremacy and hinders the ultimate performance of diamond electronics and compromise the device yield and hence manufacturing future. The presence and evolution of strain in diamond plates during epitaxial growth and device fabrication raise production cost and reliability concerns, and ultimately put the feasibility of a high-quality supply chain in question. For electronics R&D, substrates of a lateral size of at least 10×10 mm$^2$ to a square inch or better two inches are desirable to produce a device or an integrated circuit on the inner surface and away from the edges. For a long time, MPACVD diamond was a synonym of subpar quality diamond. To revolutionize modern electronics, incredible effort is being directed to finding ways to use CVD to economically grow defect-free, large area diamond thin films. To reach the described wealth of potential applications, we need a method to produce diamond as a thin film that is scalable in size and available in plentiful amounts.

In 2020, the Department of Energy (DOE) funded programs in the search of growing large diffraction-grade diamond crystals for the manufacturing of X-ray optics for next generation brilliance light sources (synchrotrons and free electron lasers). The requirements were based on some early feasibility demonstrations [32] that diamond crystals could attain sizes 6×6×4 mm$^3$ with perfect reflectivity >99% and theoretical bandwidth of 200 nrad over an area of 5×5 mm$^2$. Those essentially meant that dislocation density must be as low as <10$^0$ cm$^{-2}$. As Fig.3 highlights, it was only possible in HPHT crystals and DOE pushed for finding ways to enable new CVD means for economic production. In 2023, DARPA announced a new program to try and find pathways toward top-quality diamond produced economically for future radiofrequency (RF) and power electronics. Requirements were to grow diamond substrates greater than 50 mm with dislocation density below 10$^3$ cm$^{-2}$. In all announcement cases, such diamond would automatically guarantee superior electrical, thermal, and mechanical properties. Both DOE and DARPA announcements were eyebrow-raising but were informed by and came on the hills of new technological discoveries in MPACVD of SCD: mosaic, epitaxial layer outgrowth and flip-seed homoepitaxy and heteroepitaxy on iridium.

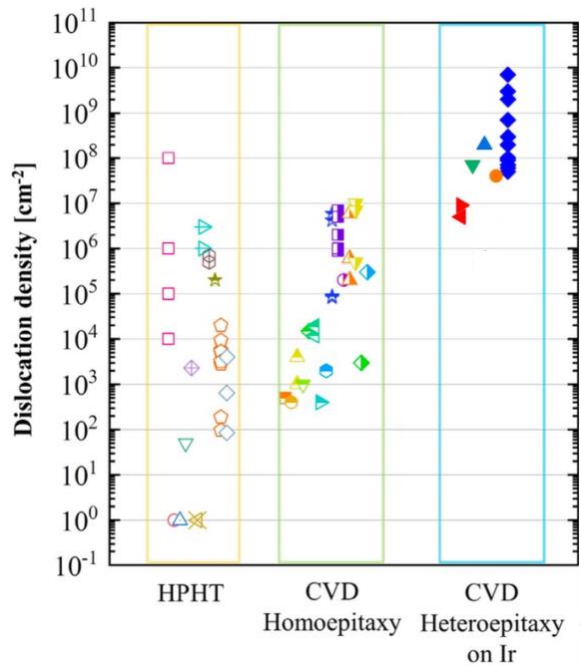

Fig.3. A comparison chart that maps out the state-of-the art quality of HPHT and homo- and hetero-epitaxially grown diamond in terms of their dislocation density. Data is taken from Ref.[33].

**Diamond CVD background**

Si and Ge are in the carbon group of the periodic table. Unlike Si and Ge, carbon can assemble in different crystal forms, so-called allotropes, as explained and example in Fig.4. There is abundance of carbon allotropes known to date [34, 35], and only one of them assembles in the form that gave the name to Si and Ge crystal structures, i.e. diamond cubic structure. Graphite is the most thermodynamically stable material (ground state) and diamond is metastable. To reverse graphite into diamond one needs high pressure and high temperature. It is impossible from thermodynamic point of view to produce diamond under low pressure conditions. The discovery patented by Eversole and all who came later unambiguously demonstrated the opposite. The phase diagram of diamond synthesis is summarized in Fig.5. In any CVD method, for SCD growth methane feedstock of CH$_4$ (3-5% of the mixture) diluted in hydrogen (95-97% of the mixture) is used. Hydrogen is a secret sauce and does all the important jobs in the MPACVD reactor.

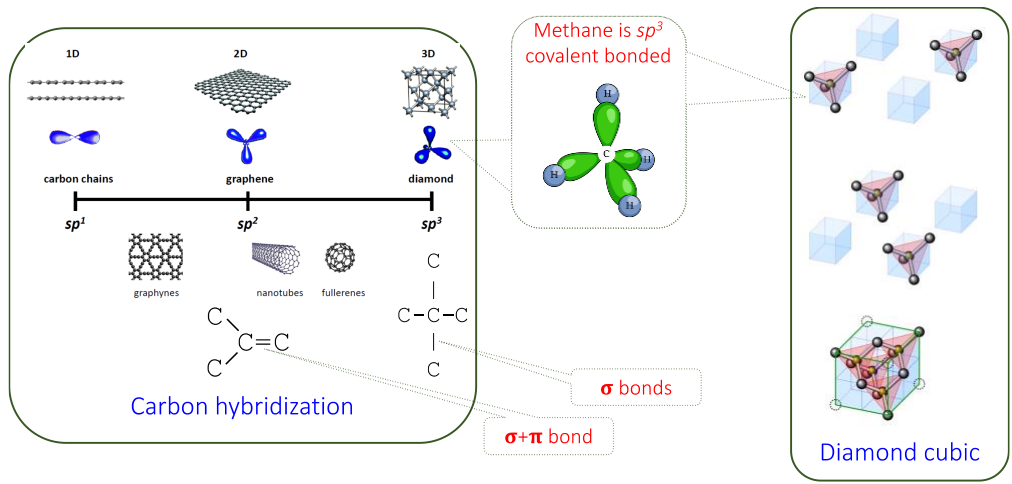

Fig.4. A conceptual diagram comparing *sp*, *sp²* and *sp³* hybridization of carbon and lists corresponding allotropes most praised in electronics.

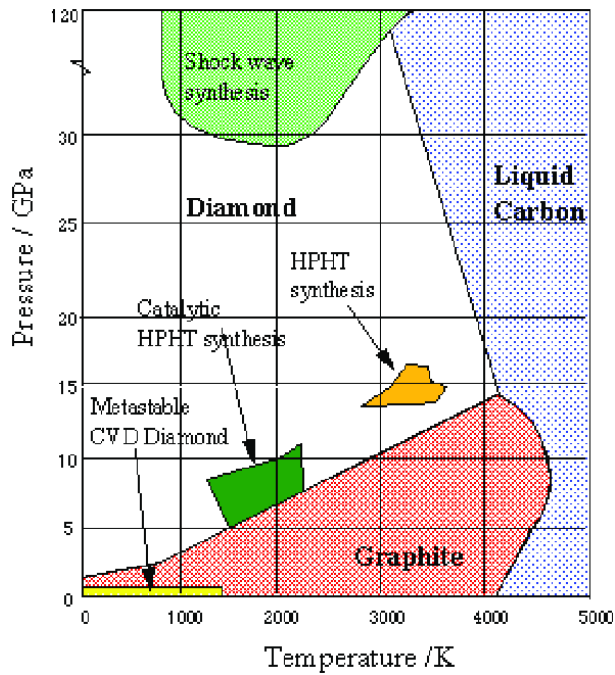

Fig.5. Diamond-graphite phase diagram.

Hydrogen defines the plasma properties, such as electron concentration and energy distribution and gas temperature. Graphite is *sp²* bonded and diamond is *sp³* bonded. The difference is that *sp³* bond must contain only single bonds, called sigma bonds, between carbon atoms. Graphite, or rather graphene, is a *sp²* planar structure giving away 2D materials with conductivity. Diamond *sp³* is a 3D structure. It actually replicates methane. By adding four methane molecule shaped unit cells, one constructs the diamond cubic crystal. This is summarized in Fig.4. In order to maintain single sigma-bonds to build the diamond cubic structure methane must be converted into methyl and remain suspended in atomic hydrogen. In early 1990's, a critically important body of work emerged that showed extremely efficient dissociation of hydrogen gas into atomic hydrogen in 1-1,000 Torr if gas temperature is 1,000-3,000 K. Fig.6a shows that it is possible to dissociate over 50% percent of molecular hydrogen. It was shown by modelling and experiment [36] that hydrogen plasma driven by a microwave source at 1-1,000 Torr can produces that hot of a gas. Upon dissociating, atomic hydrogen attacks methane and produces the reactive methyl radical, a fundamental building block of diamond growth. It appears that methyl production through a reaction $H + CH_4 = CH_3 + H_2$ is much more energy favorable (activation energy of 38.07 kJ/mol [37]) than through direct decomposition reaction $CH_4 = CH_3 + H$ (activation energy of 438.89 kJ/mol [38].) This is the first abstraction reaction. Additional relation between hydrogen atomization ratio and methyl production rate was established by Goodwin and Butler [39], see Fig.6b. It is seen that approximately 1-10% of atomized hydrogen is enough to produce maximal amount of methyl. This corresponds the standard operational regimes of MPACVD reactors of 100-200 Torr. Finally, atomic hydrogen saturates the methyl system and locks methyl radicals into single bonded configurations. One could think of diamond synthesis as a series of hydrogen abstraction and exchange reactions as summarized in Fig.7. Then methyl radical interacts with the substrate or other radicals through a series of follow up hydrogen abstractions and creates the famous diamond lattice hex where all carbon atoms are sigma bonded. Lastly, as Eversole showed, hydrogen actively and preferentially etches $sp^2$ graphitic content at >1000 K temperatures and hence additionally make diamond production net positive [40].

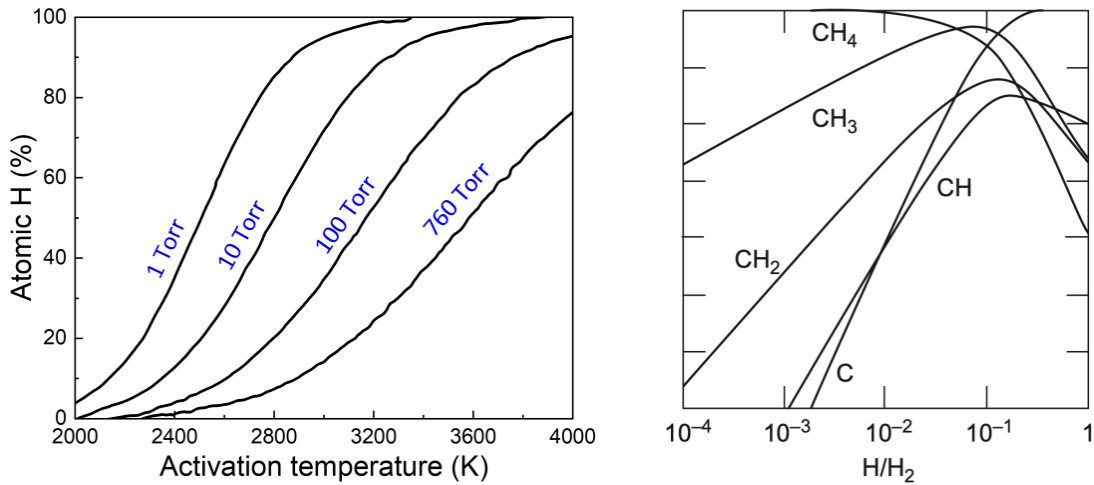

Fig.6. (a) Efficiency of atomic hydrogen generation at different gas temperatures and pressures, adapted from Ref.[41]; (b) The atomic hydrogen ratio effect on carbon species at a typical gas temperature 2,500 K, adapted from Ref.[39].

Little has changed in terms of understanding of basic diamond growth kinetics in methane/hydrogen plasmas. Main efforts and successes were achieved in finding innovative ways to overcome substrate related choke points and substrate holder re-designs for scalable productions. These successes and innovations are reviewed for the remainder of this review.

In plasma:
$$CH_4 + H = \bullet CH_3 + H_2$$
On the substrate surface:

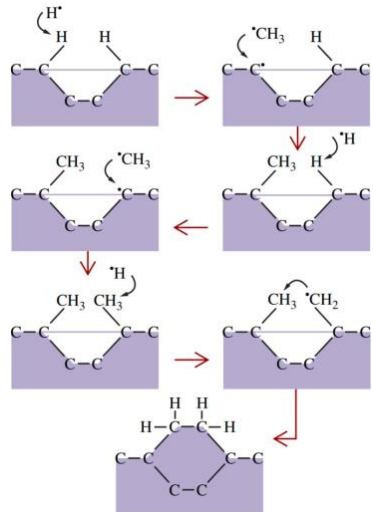

Fig.7. Hydrogen abstraction reactions drive diamond synthesis.

**State-of-the-art**

*CVD SCD lift-off and substrate recycling*

It must be noted that in 1-to-2-inch diamond plate production, separating the grown layer from the substrate using laser ablation is no longer a viable option due to the kerf loss [42]. Currently, the state-of-the-art solution is the lift-off via interface ion implantation. Ion implantation is a common technique in semiconductor industry used for doping, where dopant ions directly bombard a wafer. Dopant depth and distributions within a substrate are controlled by adjusting kinetic energy and charge state of the ions. A variation of this technique was first applied to diamond [43, 44] with a goal of generating sub-surface damage such that the sub-surface region is graphitized and the top growth diamond surface remained intact. Graphitized region is then more prone to chemical attack or dry etching, thereby opening up the possibility of detaching two sections of the post growth sample separated by the engineered graphitic interface, a process coined as lift-off. Fig.8 shows a diagram of the process. Typically, MeV carbon ions are used to irradiate diamond to engineer the graphitized interface. Because the very top surface retains its crystallinity it can be used as a seed for diamond deposition. The damaged region is further graphitized under elevated temperature during CVD growth (~900 °C). The separation is extremely robust with minimal kerf losses and achieved by different means such as hot chromic acid, thermal oxidation, and others. The lift-off technique was proven to be very effective for large plate production with no inherent size limitations and hence became a method-of-choice for recycling high quality HPHT seeds, that are very costly, hence for overcoming the supply bottleneck for scalable SCD production in MPACVD at lower cost.

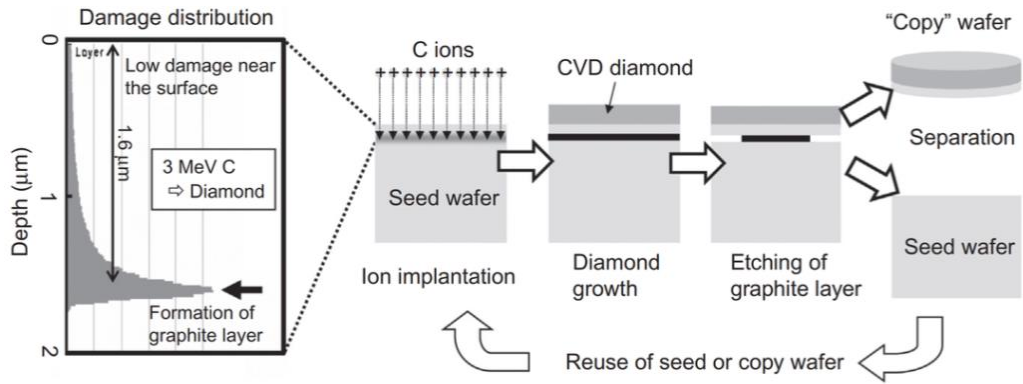

Fig.8. The concept of ion implantation enabled lift-off. Sub-surface damage distribution within the diamond seed is shown when irradiated with MeV C ions. A step-wise schematic illustrating the wafer replication process by recycling the substrate seed [45] is shown.

*Substrate pocket holder design and polycrystalline diamond rim termination*

In MPACVD, the parameters for a geometric growth progression are expected to remain the same with time once the macroscopic reactor parameters like the gas flow, pressure and microwave power are set. It appears to be an incorrect assumption. As the CVD crystal grows thicker on the substrate, gaining thickness similar to that of its substrate, additional effects caused by sample-plasma interactions start playing role. A practical method allowing for quasi-stable growth conditions was proposed and realized at AIST in Japan. They introduced a so-called enclosed (or pocket) type molybdenum holder in 2005 [46]. The design and the difference against the traditional open type of holder are illustrated in Fig.9. In the original paper, it was demonstrated that growth condition stability was achieved. The pocket design resulted in a striking improvement of the grown crystal quality as long as the sample remained recessed in the pocket.

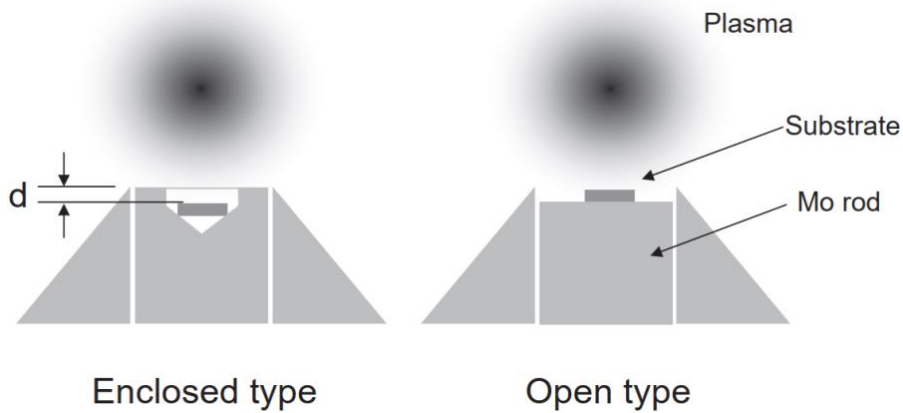

Fig.9. Schematic illustration of the enclosed/pocket and open type sample holders: the sample is recessed in the pocket to distance $d$ [46].

At MSU, Asmussen's group [47] found that there is an additional benefit of the pocket holder design. It helps mitigating the polycrystalline diamond (PCD) rim formation. Until then, PCD rim formation was yet another shortcoming in the CVD growth. PCD formation is a net negative effect resulting in reduced lateral area of the grown SCD films. Fig.10 illustrates the concept of the discovered mitigation. The general understanding of this effect is that plasma does not couple to the substrate edges, thereby not causing overheating at the edges. As a result, it does not change the growth conditions from being favorable to grow SCD into PCD, which contains a large fraction of bulk defects like graphitic grain boundaries. Such a simple modification of the holder was proved to be absolutely indispensable in producing higher quality crystals of simultaneously larger lateral area.

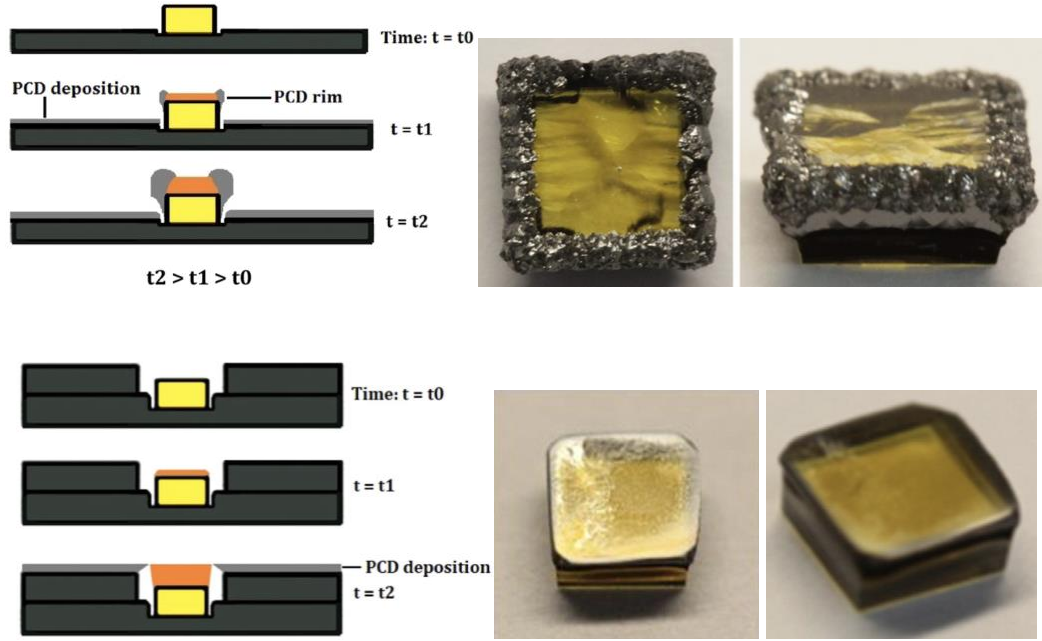

Fig.10. Open (top row) versus pocket (bottom row) holder comparisons regarding the PCD rim formation: Cross-sectional views of SCD synthesis versus time and optical images of as-grown crystals. As SCD growth versus time proceeds the new SCD grows without a PCD rim both vertically and horizontally. PCD is only deposited on the top of the substrate holder. The yellow color denotes the HPHT SCD seed substrate, orange denotes the synthesized CVD SCD and the gray area identifies the deposited PCD material.

*Mosaic technique*

This technique should not come as a surprise if one remembers how thin solid films grow undergoing through nucleation, isolated crystal/island growth and coalescence, as illustrated in Fig.10. The first demonstration of the mosaic technique was experimentally reported in 1994 at MIT Lincoln Laboratory [48]. Several hundred of cubic micro tiles, defined here as diamond cubes with sides no larger than 300 μm (Fig.11) were used. The mosaic seeds were [100] HPHT crystals. Crystal tile sizes and shapes were sorted, assembled and kept together in a specialty holder on a silicon wafer as illustrated in Fig.11. Approximately 20 μm of homoepitaxial diamond was grown over this substrate using a plasma CVD process, forming a continuous layer over the mosaic substrate as illustrated in Fig.11. The resulting layer was highly defective, but it validated the basic concept of mosaic growth in demonstrating that a continuous homoepitaxial layer can be achieved even with multiple interfaces and imperfect alignment between tiles. The study also showed how quantitative evaluation of relative misalignments is straightforward, regardless of the number of tiles. After this trailblazing demonstration, the mosaic approach has been accelerating and deepening as a branch of diamond technology with a few technological milestones, including a 3D mosaic crystal [49].

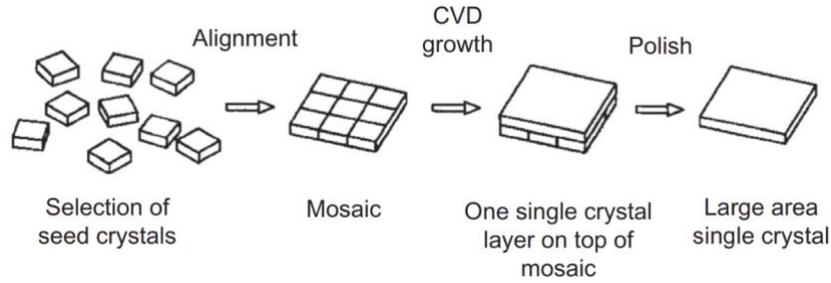

Fig.10. The principal concept of achieving large area diamond substrate through mosaic technique.

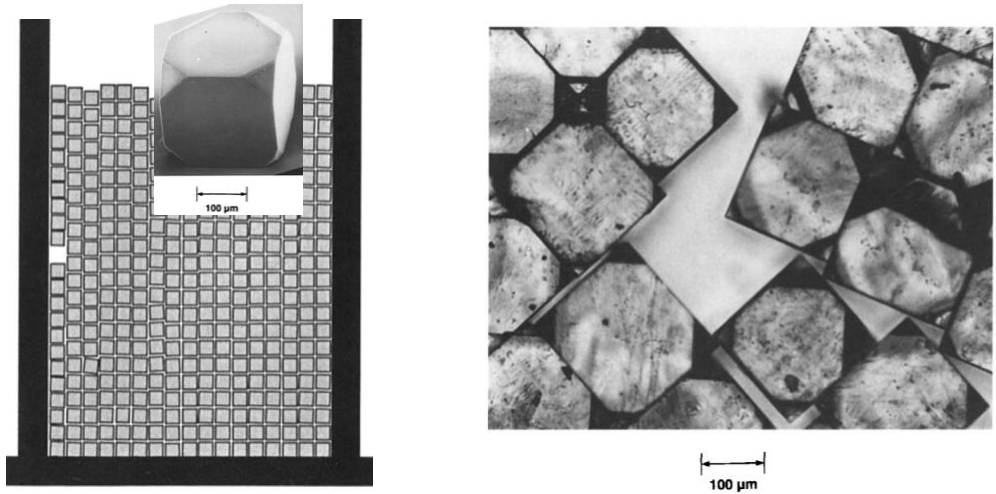

Fig.11. (left). Top view of schematic drawing of assembly of cubic crystals in a square frame. Inset is a scanning electron micrograph of a single cubic diamond crystal, where the square faces of the cube are {100} planes and the corners are faced with {111} planes. (right) Optical micrograph showing joined mosaics.

As Figs.10 and 11 illustrates, conceptually the mosaic approach is straightforward. From practical standpoint, isolated tile assembly remained a challenge because the height distribution between individual tiles and their relative in-plane (twist) and out-of-plane rotations (tilt and torsion) (see Fig.12 for explanation) and interface gaps were later found critical for successful growths. The main requirements being creating conditions were stress at the interface is 0.1-0.3 GPa [50, 51] guaranteeing no cracking of the wafers upon thermocycling in the reactor and during post-growth device/IC fabrication steps.

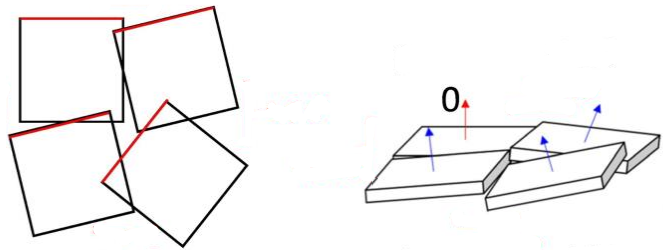

Fig.12. Four tiles placed together illustrating the TTT concept of twist (left) and tilt/torsion (right) angle misorientations [52].

Stringent requirements on tile preparation and assembly in terms of tile height, twist, tilt and torsion and limited substrate sourcing slowed down the progress toward attaining high quality SCD plates of 10+ mm in linear size. To make the mosaic technology economic, multiple sourced high quality substrate tile had to be found, characterized by X-ray diffraction (XRD), laser cut, polished, remeasured by XRD making sure the correct crystallographic qualities. There was slow progress until a cloning technique was invented at AIST, Japan [53]. The group realized that if the ion implantation and lift-off process is repeated using the same seed, each produced plate would have identical crystal orientations and other geometrical features. The group proceeded with developing mosaic technique by defining these produced plates as clones. The cloning process starts by ion implanting a high purity HPHT Ib-type seed. Each clone and following mosaic plates are all grown by MPACVD and cloned through the lift-off (ion implantation plus etching). Conceptually, the full process is illustrated in Fig.13.

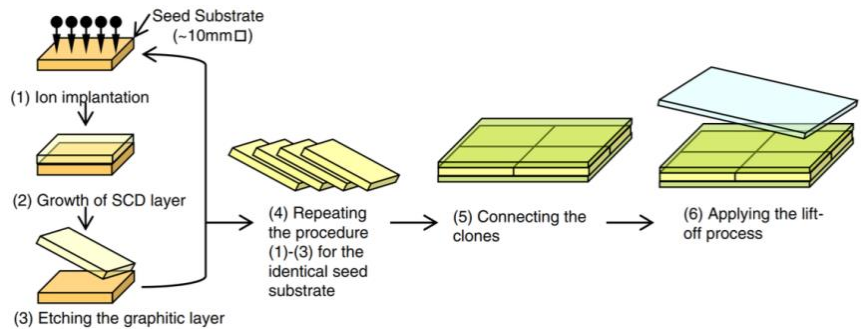

Fig.13. The concept of production of cloned tiles by ion implantation, and of the resulting large plate production [54].

Each cloned tile is fine-tuned by laser cutting and polishing leading to the resulting angle misalignments from the mosaic re-assembling process of about 2° which yielded excellent results and effectively showed a path for solving the substrate sourcing bottleneck. As Fig.14 highlights, the process was successful at the 1-inch scale when four tiles were joined. Boundaries above the original tile interfaces are not visible optically and are labeled with arrows in the figure.

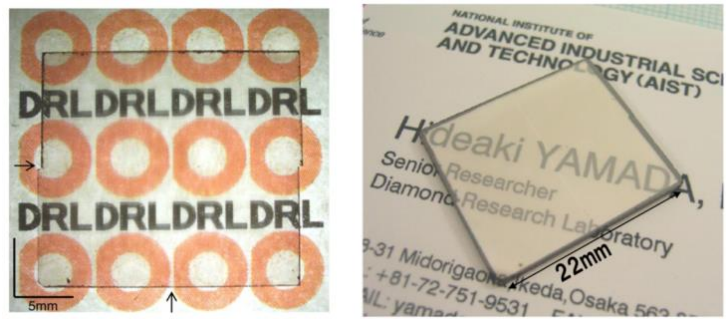

Fig.14. Polished mosaic plate 22 mm wide as produced by the tile cloning mosaic method based on four tiles put together as a substrate.

If the cloning process is repeated over and over again, the resulting diamond plate area is amplified exponentially. The same group expanded the total area of CVD grown SCD plates first to 20×40 mm$^2$ and later attempted a 2 inch wafer, a critical benchmark size for semiconductor industry, and successfully produced a 40×60 mm$^2$ diamond substrate by joining together twenty-four 10×10 mm$^2$ cloned SCD plates [55]. The thickness of the plate was 1.8 mm. These two milestone results, and one other one reported in Ref.[56] are shown in Fig.15. These 2-inch results hold the record and show the potency of the mosaic technique for revolutionizing diamond electronics.

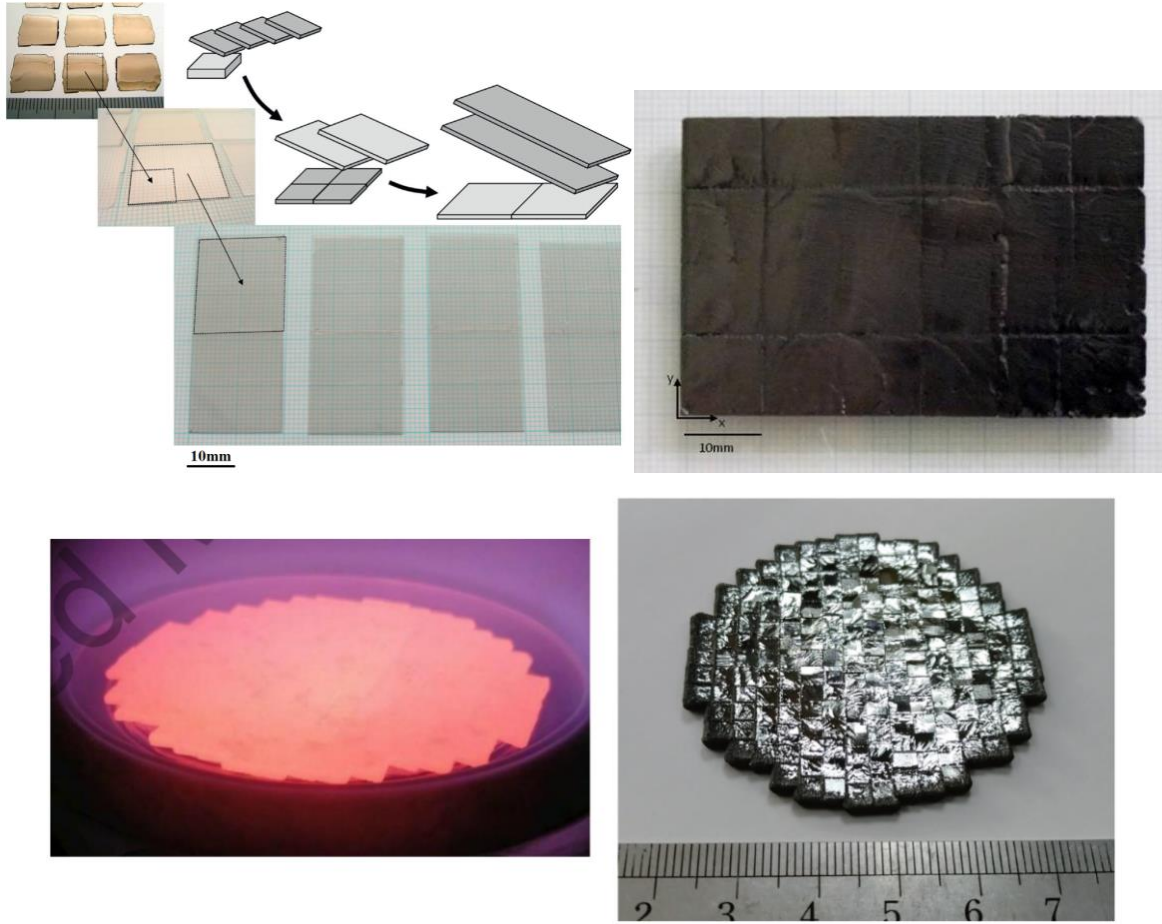

Fig.15. (top) Demonstrated cloning processes resulted in 20×40 mm$^2$ and 40×60 mm$^2$ SCD of mm thicknesses; (bottom) Two-inch diameter SCD during growth and as grown. In this case, thickness non-uniformity, center to edge, was found to be as high as 25%.

Recent work by Diaz at MSU [57] revisited the mosaic growth and created a systematic list of critical requirements for successful 1 to 2 inch growths as

First, it was practically found that the gaps between the tiles, to be successfully bridged and overgrown in the homoepitaxy, have to be less than 50 μm [53]. Larger gaps do not merge.

Second, as best described by Findeling-Dufour *et al.* [58, 59] and supported much later by Wang *et al.* [51], the height steps between the tiles (resulting from the tile thickness differences) are favorable for tile joining as it facilitates so-called step flow growth and bridging per Fig.16. Without this condition the mosaic growth fails. Whether there is a critical height step exists remains unknown.

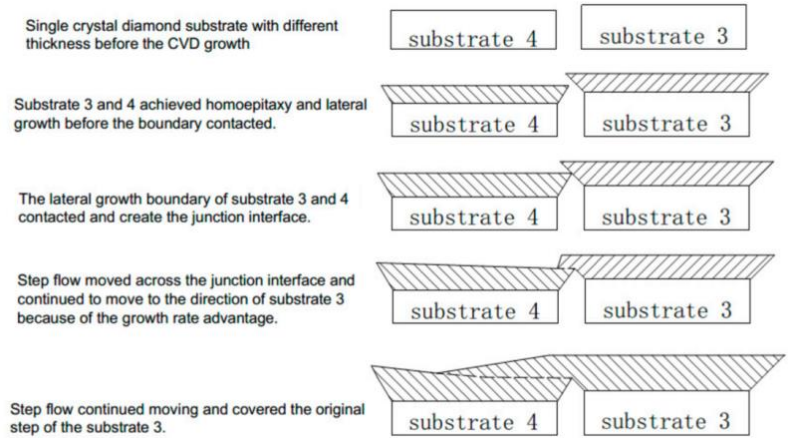

Fig.16. Proposed mechanism of junction interface formation under the surface step flow in CVD mosaic growth. Adapted from Ref.[51].

Third, relative in-plane tile orientation (twist angle as in Fig.12) matters. Because in high quality homoepitaxy orientation of the substrate tile is conserved, the discrepancy between step flow growth direction and actual orientation, as set by the twist angle, obviously must lead to internal stresses [50, 51] often causing cracking. The larger the twist the larger the stress. The one case where the four tiles held together without cracking were assembled with relatively small misorientations [52]. This is why the cloning technique was so successful at increasing substrate areas substantially as it virtually eliminated misorientations, twist, tilt and torsion.

Fourth, to minimize internal stresses as the mosaic crystal assembles growth conditions must be incentivized to allow for the resulting boundary to form hundreds of microns away from the actual touching boundaries (see Fig.16). To facilitate this process, lattice off-cuts are introduced, where the cut-off means a few degrees difference from the perfect cubic crystallographic angle. The resulting step flow growth leads then to a boundary shift thereby improving the boundary crystal quality.

*Epitaxial layer outgrowth (ELO) technique*

At Michigan State University, there were a series of important findings [60, 61] that resulted from the intertwined effects of pocket holder and step flow growth, as pictured in Figs.17. Because the parasitic effect of the PCD rim formation is suppressed, single crystalline outflow/outgrowth is allowed. The CVD crystal literally gains the lateral dimension as it gets thicker with time. The original papers found that it was possible to double the lateral area without any additional efforts! This area increasing effect was dubbed as epitaxial layer outgrowth or ELO. Further, it was discovered that there is an asymmetry effect that may be induced if the diamond substrate is off set in the pocket holder as highlighted by Fig.17.

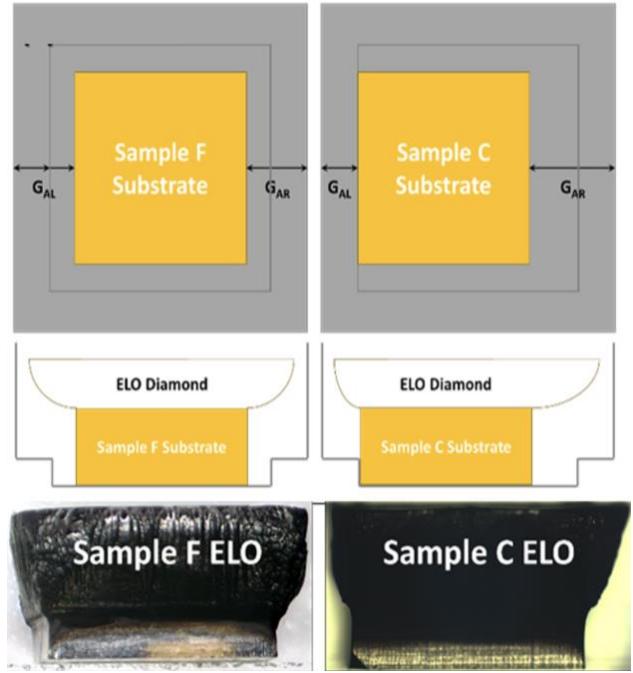

Fig.17. Schematic and results of symmetric and asymmetric ELO in the pocket during CVD.

Our recent work at MSU [62] extended understanding of the asymmetry effect and overall ELO kinetics by analyzing the growth results in the framework of the Avrami model [63] describing the gas-to-solid phase transformation through temporal lateral gain function *L(t)* as

$$L(t) = A(1 - e^{-t/\tau}), \quad (1)$$

when boundary conditions are set for the simple straight box pocket holder design (Fig.18). Here *A* is the max lateral gain (measured in mm) and $\tau$ is the characteristic time (measured in hours.) Eq.1 is a trivial case of the first order exponential decay growth. The lateral gain rate slows down exponentially as the distance *G* between the growing CVD crystal and the pocket side wall reduces. While trivial, Eq.1 perfectly describes the resulting growths in both symmetric and asymmetric cases.

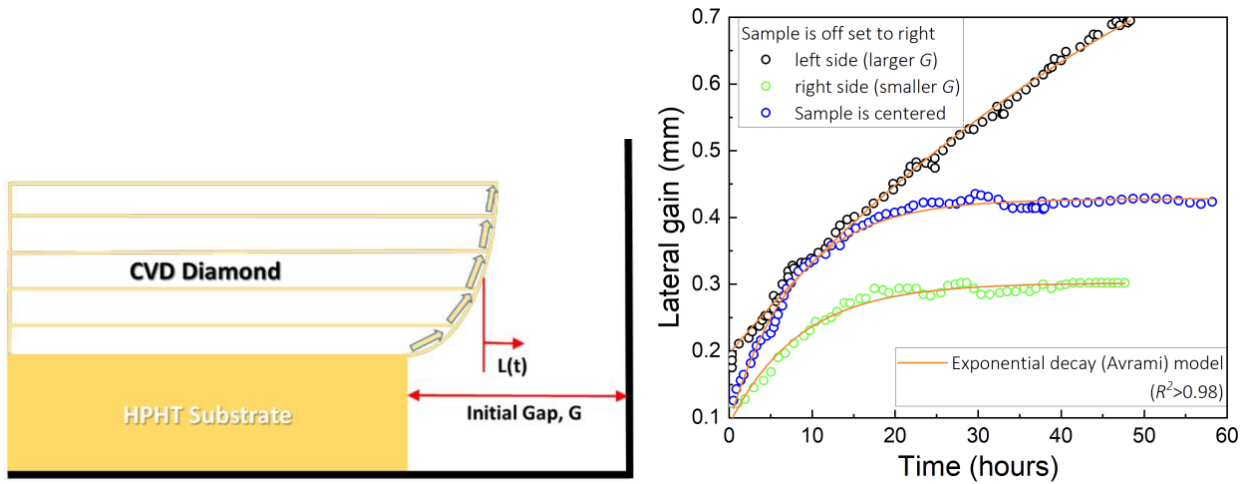

Fig.18. (left) A step-like schematic of the ELO process resulting in exponential decay mode; (right) Lateral gain experimental data fit by Eq.1.

The next question that arises is whether the holder-SCD system can be manipulated such that the growth rate $dL(t)/dt$ is a constant. Intuitively, it is clear that the holder needs to be redesigned such that the gap $G$ between the growing CVD crystal and the pocket wall is kept constant. This then mitigates the exponential decay and allows for the desired constant growth rate. The geometrical means to achieve and control the optimal constant gap distance is by making an angled pocket holder. Again, it is intuitively clear that the angle has to be between 0 (open holder design causing unwanted PCD rim) and 90 (square box design causing exponential decayed ELO) degrees. This concept is illustrated in Fig.19. A series of holders were created where the most promising were the ones with an angle close to 45 degrees. Fig.20 compares results obtained for the holders that have opening angle varied with respect to 45 degrees; it compares 60- and 37-degree pocket holders. Because the pockets now are semi-open, some PCD deposition is expected to form that would cause the gap to close from 37 degrees toward 45 degrees thereby inducing quasi-linear ELO. The inset in Fig.20 shows the cross-section of a sample typically grown in the 37 degree design showing the straight line crystal expansion directly confirming that linear ELO mode was achieved. Through additional optimization steps, it was shown that the linear size of the homoepitaxially grown CVD crystals can be doubled and the area can be quadrupled which is a next significant step forward in grown truly single crystal diamond without internal interfaces like it is in the mosaic approach.

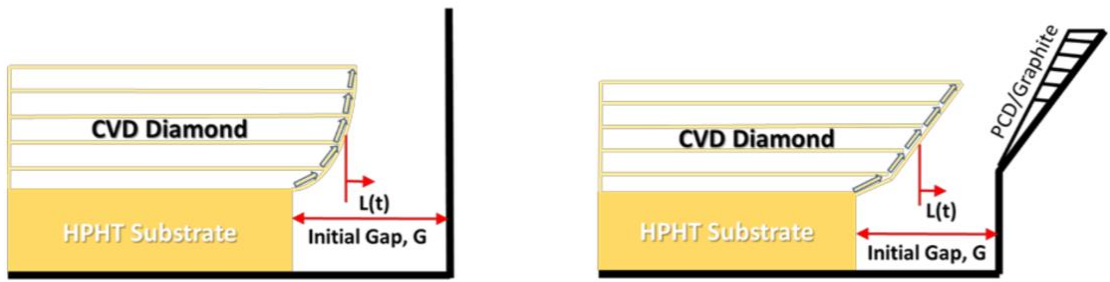

Fig.19. Linear lateral growth can be achieved by using properly angled pocket holder.

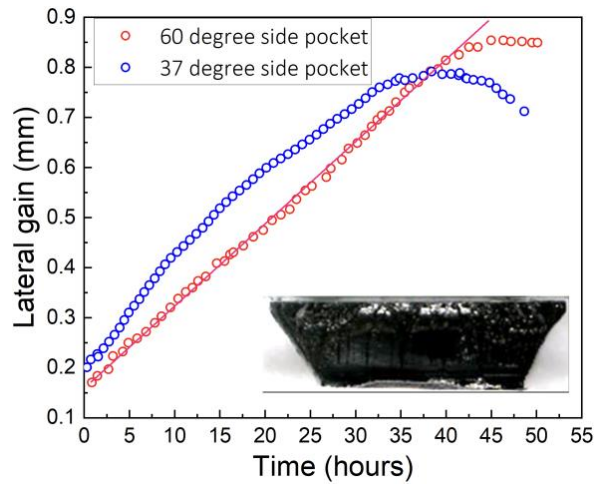

Fig.20. Comparison between 60 and 37 degree pockets leading correspondingly to sublinear (blue symbols) and linear (read symbols) ELO profiles during CVD.

*Heteroepitaxy*

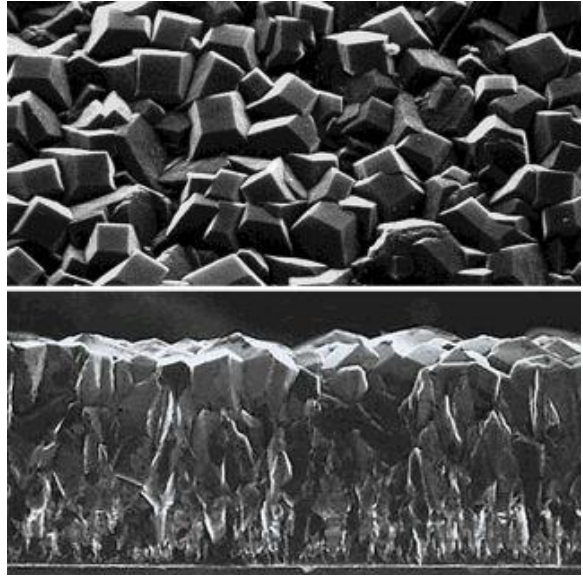

Fig.21. Micrographs of top surface view (top) and the cross section (bottom) of a polycrystalline diamond film [64].

The growth of single crystalline diamond on non-diamond substrates has long been desired and attempted. If achieved, it would give new opportunities where diamond CVD is no longer dependent on limited supply HPHT substrates. One prime contender was the use of silicon since it has the diamond cubic lattice, albeit a different lattice constant. Additionally, silicon has been the established semiconductor material for the last 50 years and practically flawless wafers 300 mm in diameter are easily available. While there is a plerophory of data and research growing diamond on silicon, any attempts to grow single crystalline diamond were futile. Any growth of diamond directly on silicon, whether it be on unseeded wafers [65] or when applying improved seeding conditions (diamond nano particles [66], dip coating [67] or surface carburization [68]) result in the formation of polycrystalline diamond (PCD) instead. The major problem will all of these grown PCD films is that crystal nucleation occurs in random orientation. Fig.21 shows a cross section of such an exemplarity PCD growth. Thus, even though individual crystals coalesce when growing the film thicker, the existence of the random orientation prevents the ultimate coalescence into a single crystal, leaving behind grain boundaries. As in mosaic technique (Fig.12), the TTT misorientations in between individual crystals limit how much individual crystals can combine. In order to successfully grow SCD onto silicon material, it would be necessary to overcome this fundamental challenge, as in have all nucleation sides aligned with respect to twist and tilt. The most promising technique explored is bias-enhanced nucleation (BEN). BEN has been demonstrated to produce highly oriented nucleation sides on silicon. Subsequently, these nucleation sides can be coalesced effectively and grow into highly oriented diamond. The effectiveness of BEN in creating highly oriented diamond has been demonstrated on silicon [69-71] as well as on silicon carbide [72], another wide bandgap material that is of high interest for heterointegration with diamond. Fig.22 shows the top surface of such a highly oriented diamond film on silicon. It can be clearly seen, that the orientation of the diamond is almost exclusively alongside the (100) orientation. The resulting film appears more platelet like (something one could call assembled micro-tiles with minimized twist, tilt and torsion) than what randomly oriented PCD looks like (Fig.21). Nevertheless, remaining substantial TTT variance is still visible. While BEN offered the most promising path in achieving direct heteroepitaxy of SCD on silicon, the process ultimately fell short.

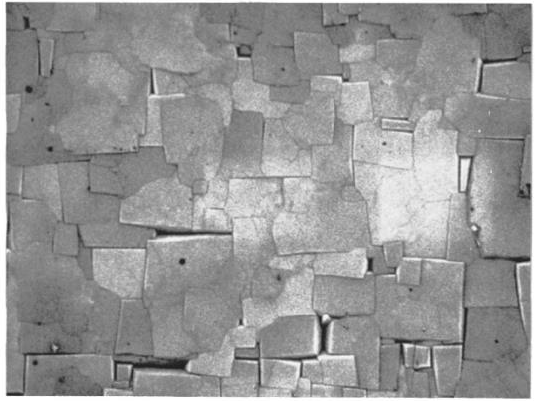

Fig.22. Micrograph of a highly oriented heteroepitaxially nucleated diamond film on (001) oriented silicon wafer [71]. Scale bar is 10 microns.

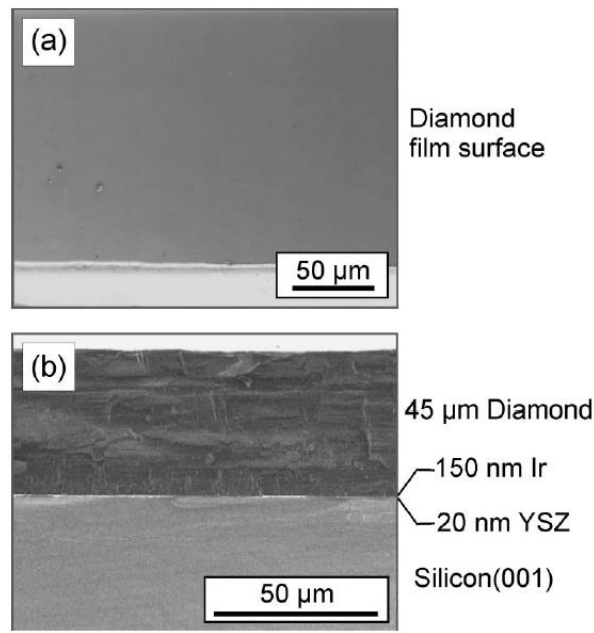

Fig.23. Micrograph of a 45 µm thick heteroepitaxial diamond film deposited on Ir/YSZ/Si(001): (a) diamond film surface; (b) cross-section [73].

In the end, iridium came as a remedy and proved to be the only suitable material, where heteroepitaxy of SCD has been demonstrated. Since its first successful reporting by Schreck et al. [74] every scientific report on SCD growth heteroepitaxy is done on iridium, albeit at a variance of substrate materials. Fig.23 shows the top growth surface achieved in a later publication by Gsell et al. [73]. It can be clearly seen that the grown diamond film is completely smooth and one continuous film. There is no resemblance of individual grains. The key reason that SCD heteroepitaxy on iridium succeeded, while it failed on silicon is that the lattice mismatch between diamond and iridium is only 7.1%, while it is 35% between diamond and silicon. More recently, in 2017 Schreck et al. [75] also introduced the ion bombardment induced buried lateral growth (IBI-BLG) mechanism, which explains how heteroepitaxial growth takes place. There, ion bombardment from the plasma induces the formation and lateral spread of epitaxial diamond within a narrow (~1 nm) carbon layer. This allows for the spread of buried epitaxial islands, created by original nucleation sides, over distances of several microns. During this process, secondary nuclei are being created continuously in the vicinity, which in turn cause the continuous lateral overgrowth with diamond. So far, this process is unique to growth of diamond on iridium surfaces.

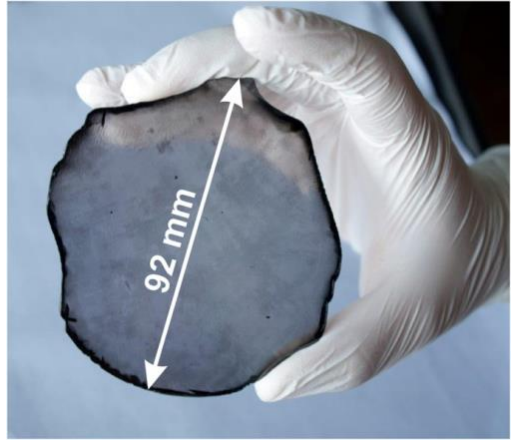

Fig.24. Optical image of a 92 mm diameter SCD grown on Ir via heteroepitaxy. The wafer has a total thickness of 1.60±0.25 mm and its weight is 155 carat [75].

Once SCD heteroepitaxy has been successfully developed on iridium, it allowed to synthesize freestanding SCD wafers close to 100 mm. See Fig.24 for such an example of a 92 mm (3+ inches!) SCD wafer grown via heteroepitaxy. However, the process faces some unresolved challenges. For one, the deposition of iridium onto suitable carrier wafer materials is non-trivial. For example, in order to deposit iridium onto silicon, an yttra-stabilized zirconia (YSZ) [73] or strontium titanite [76] interlayer is necessary and iridium deposition temperatures need to be above 800 °C. YSZ turned out to be the favorable choice since it reduces both, the in-plane twist of the nucleation centers and also the out-of-plane tilt, while strontium titanite only reduces twist. Other suitable carrier wafer materials are sapphire [77] and magnesium oxide [78] but these substrates contain their own individual additional challenges. Another one is as follows. Because diamond deposition takes place at substrate temperatures >800 °C, any of stack substrates used for growth, for example diamond/Ir/YSZ/silicon, have significantly different individual thermal expansion coefficients, thereby causing significant thermal stresses into the material stack and, as a result, often leading to cracking of the resulting SCD wafer. Process optimizations attempting to minimize the amount of stress formation are still ongoing [79, 80]. For example, processes using magnesium oxide now use the KENZAN technique, which provides natural delamination of the heteroepitaxially grown SCD wafer. This prevents wafer cracking and helps substantially reduce threading dislocations [81].

*Flipped seed technique*

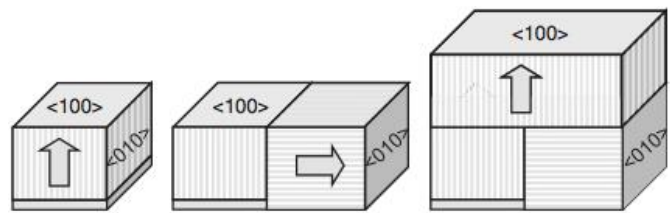

Fig.25. Schematic illustration of flipped crystal growth on three equivalent <100> orientations [82].

The approach of growing on various sides of a crystal in order to expand its lateral dimensions was first introduced by Mokuno et al. [83] as "side-surface growth", but was also known as "flipped side" or "flipped crystal growth". The technique employs the fact that all {100} crystal orientations are equivalent, and hence allow for equal growth whether it occurs in <100>, <010> or <001> direction. Early reports on the flipped crystal approach altered the growth between the [001] and [010] direction; it is illustrated in Fig.25, where three subsequent growth iterations were performed altering between [100], [010] and again [100] growth. Each respective cube represents one growth step. The small lines indicate the individual growth directions of each growth iteration. In between growth steps, the diamond would typically be laser trimmed and polished to remove any PCD or otherwise parasitic growth that had occurred and would negatively impact future growth iterations. In Fig.25, the original dimensions were doubled and, as result, the area quadrupled. Fig.26 shows a photograph of a diamond grown by the flipped seed method. Surface enlargements up to $10\times10$ mm$^2$ were reported using this technique [84]. A slight variation of the flipped crystal approach originating from the same research group was introduced [83]. There, growth iterations on three adjacent sides were performed. This variation leveraged the equivalence of the {100} sides. Diamond wafers half inch in size have been demonstrated [83].

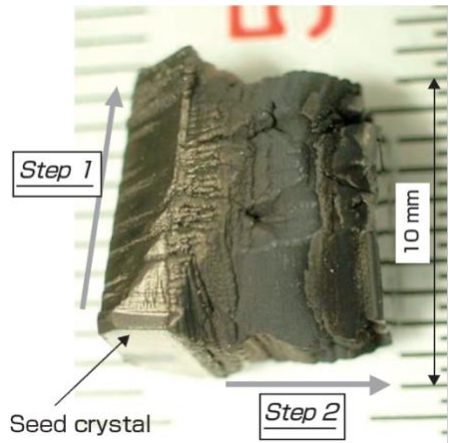

Fig.26. Example of lateral growth using three iterations of the flipped seed method [84].

The two biggest problems associated with using flipped crystal growth is that defect density and the local stress increased significantly at the interfaces, where the actual crystal flipping occurred. This was experimentally verified by Mokuno et al. during their half-inch wafer demonstration [83]. Fig.27 shows the photograph of the laser cleaved and polished surface of a diamond that was enlarged to a half-inch. Fig.27a shows a half-inch SCD wafer that was subsequently grown and separated from the surface of the image shown in Fig.27b. Fig.27c shows the birefringence pattern inside the wafer Fig.27b highlighting the regions of strain and defect agglomeration, which transferred from the original enlarged crystal (Fig.27a) into the newly grown half-inch wafer (Fig.27b).

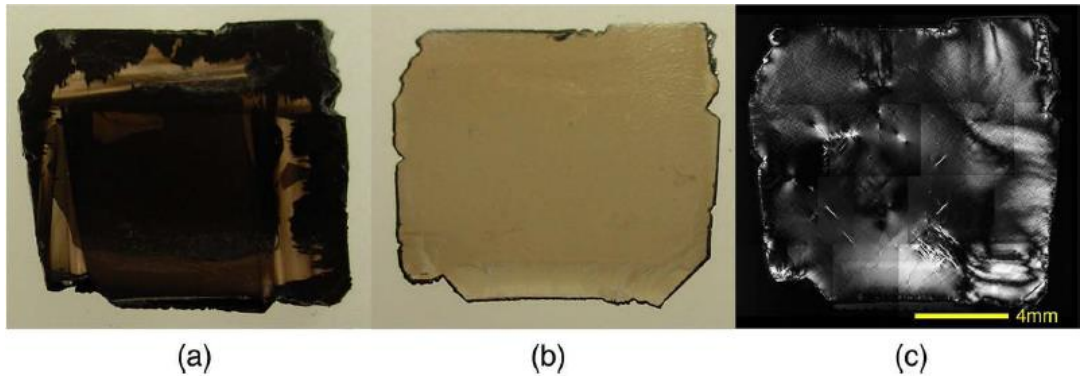

Fig.27. (a) Photograph of a half-inch seed plate obtained by cutting and polishing the diamond that was enlarged by flipped crystal growth. (b) Photograph of a half-inch SCD wafer grown from an enlarged seed. (c) Polarized light microscope image of the half-inch SCD wafer highlighting the regions of internal crystal stress.

Yet another alteration of the flipped seed method is the flipped side approach, which was proposed by Muehle in his PhD thesis [85]. While the main principle still relies on the growth of various {100} directions it also leverages the fact that defects propagate preferentially along the original growth direction [86, 87]. Friel et al. [87] demonstrated a vast reduction of the defect density when the growth direction of SCD is flipped orthogonally across different {100} directions. This was demonstrated by X-Ray topography, where defect propagation is visible as dark streaks. Fig.28 illustrates a clear separation between the [001] and [100] growth, and that the defect density in the second [100] growth is immensely reduced.

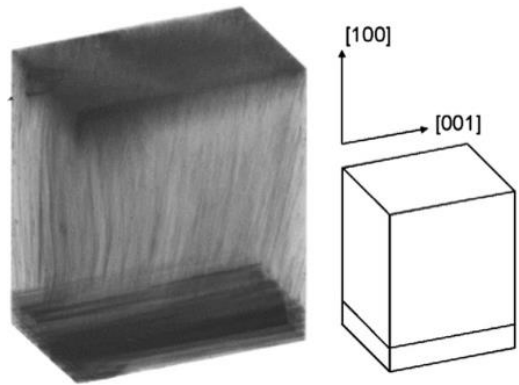

Fig.28. {111} Propagation topograph of a single crystal CVD diamond sample containing two consecutive CVD growth runs along the orthogonal [001] and [110] directions.

The goal of the flipped side approach is to solely use single directional growths as starting point for the next iteration of flipping. This is achieved by first creating a seed that is grown on its side for initial reduction of the defect density. From there, lateral enhancement is achieved by performing growth iterations to a desired thickness, then create a new flipped side crystal and repeat the process. The process for even larger lateral area expansion, starting at 13 mm node to 50 mm (de facto two inches) node, is shown in Fig.29. In the process flow, an optional step was added in the beginning which creates a 13 mm flipped side wafer for further defect reduction at the beginning of the process.

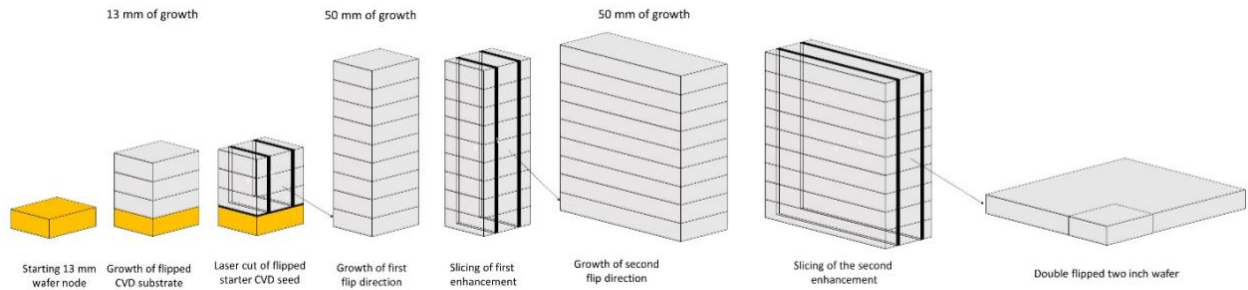

Fig.29. Schematic of the flipped side approach. Starting from a 13 mm wafer node, SCD enlargement to 50 mm is achieved via two iterations of side flipping.

While this approach sounds logical and promising, it also has major drawbacks. When doing the mosaic approach, the later enlargement of the wafer dimension is simply that of the beginning tiles. A simple vertical growth step of as little as 1-2 mm is sufficient to create the enlarged wafer, and relative area enlargements can be as much as 8× or 16×. Additionally, any further size enlargements can leverage the previously enlarged wafers. Contrary, during flipped side growth, the relation between the area to growth thickness results for each mm of growth to a similar enhancement of usable area. As a result, to achieve a two-inch wafer, a total of 100 mm of growth thickness, executed in separate enlargement steps for each of the two lateral dimensions, step is needed. Strikingly, when further size enlargement, e.g. three or four inch wafer node, are targeted the process has to start from the beginning and it does not scale from any prior progress. Any subsequent wafer enlargement comes with ever increasing requirements on total vertical growth. Another major concern for this technique is the well-known occurrence of shrinking SCD surfaces throughout the vertical growth process, especially when performing consecutive growth processes that are producing the vertical growth needed to enable the area enlargement [46]. Careful design of appropriate holders and sample placement (see Fig.20) can likely alleviate that problem, but appropriately designing optimized pocket holders that can accommodate crystals with length increasing to 2 inches or even more, and controlling the resulting growth process will amplify the complexity of an already complex and challenging engineering problem.

**Future diamond manufacturing**

This section presents the use of artificial intelligence (AI) as an emerging technology that, in the authors' opinion, has so far been underutilized in the realm of materials development. As such it has the potential to define how diamond wafers and application devices are being manufactured. Further, these techniques have the protentional to disrupt the semiconductor industry, as well as adjacent crystal growth applications and additive manufacturing.

The use of AI has pioneered and transformed, even created, many new technologies and application fields. In turn, these had an invaluable impact on our economy and society. Only 15 years after the introduction of the first iPhone in 2007, people are using a swath of apps that are built into every major operating system to instantly translate foreign language text, use biometric facial and fingerprint recognition as a safety feature, identify nearly any object in a taken picture, or use external apps for recognition of your favorite music, or birdsong. All of these apps, and many more, that greatly improving our quality of life, are fueled by clever AI [88]. AI-based content recommendations on social media apps can be seen as a blessing or a curse. Either way, its transformative impact on societies across the globe cannot be denied. The majority of these AI-fueled disruptions fall under the category associated with "tech companies". FAANG (Facebook [now Meta Platforms], Amazon, Apple, Netflix and Google [now Alphabet]) is another synonym. In a way it highlights how these five companies in particular were on the forefront of digital innovation, all of which can be tracked down to clever use of deep learning (DL), which is a subset of AI.

The takeaway is that while AI has disrupted society and large parts of the economy, similar impacts have not been seen in hardware driven technology development. So far, AI has only seen preliminary use in these, such as predictive maintenance decision making, creation and analysis of digital twins, and AI-enabled generative design creations for resource efficiency and waste reduction [89]. However, while these adoptions hint at the potential of use of AI in "old tech", it has not radically shifted the way business and R&D is conducted. One obvious holdback is the fact that development and integration of AI into complex, and often regulated hardware, is more complicated than to apply AI to digital products. As a result, materials development is still largely following a classical paradigm where human operators design new experiments based on information from post-process analysis and based on classical feedback of information. Recently, first disruptions of this concept were achieved by using AI to predict and control elastic strain in semiconductor crystalline materials, and to then experimentally demonstrate the the AI-predicted behavior [90].

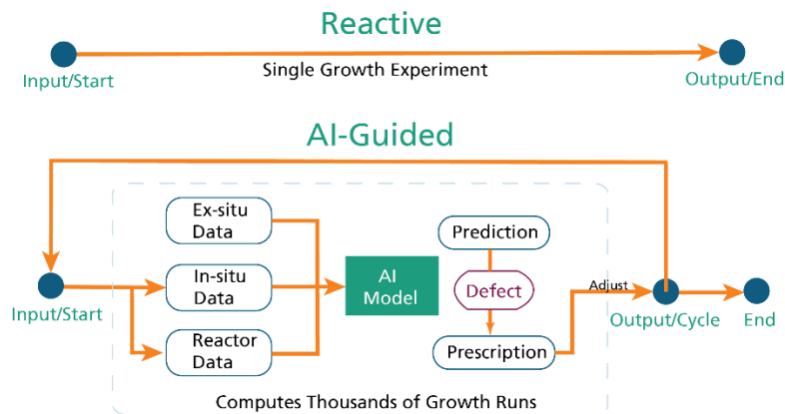

Fig.30. Schematic of a reactive (top) and AI-guide (bottom) material process development cycle.

This glimpse of how AI can transform material science and engineering inspired a reconsideration of the classical approach in experimental design. Instead, a data fed-forward guided approach was introduced and explored. The goal was to use AI in order to short-circuit the development cycle by leveraging the vast amounts of data generated during materials growth (*in situ*), which are typically underutilized in the experimental design process (Fig.30). Such an AI-guided approach inserts AI models between the start and end of the process to leverage the large amounts of data generated during synthesis. AI models trained on data distributions from *in situ*, reactor, and *ex situ* data can achieve accurate predictions of future material states, detect the formation of defects in those states, and be used as a closed-loop control. This generates a new cycle within a single process. This approach is expected to make process development more efficient by reducing the number of experiments needed to achieve a result, e.g. fewer defects in a material. Properly trained AI pipelines will self-learn over fewer experiments than an exhaustive reactive approach and can continually be updated as conditions inevitably drift during the process cycle. Such an AI-guided approach is not possible with human operators, as even experienced experts cannot, *in situ*, correlate input data with microscopic defect states and predict their formation.

Diamond is a perfect material for adoption of such an approach. For one, diamond as a material is still in its infancy when it comes to material maturation. The last 30 years of diamond research have shown that development of diamond as material, and its derived applications, follows similar timelines and cost as those of other semiconductor materials, such as SiC and GaN. For example, following the traditional reactive approach to material manufacturing development, 20 years and $10 billion dollars in R&D investment were required to introduce SiC devices to commercial markets in Japan alone (Fig.31). Despite this effort costs of SiC devices remained 3× those of Si as of 2017 due to remaining defects in the grown material [91, 92]. While SiC has seen a massive rollout in recent years due to increase demand in electric

mobility solutions, that 3× cost premium has not changed [93], and is being attributed to a still overall modest market size. It is a reasonable assessment that diamond will fall under similar timelines and cost barring disruptive ways of how material maturation is handled. Use of an AI-guide material development cycle has the potential to overcome this. When successful, it does not only vastly improve the timeline for diamond wafers and applications, but the AI-guided approach can also be applied to other crystalline materials as well as for future advances in additive manufacturing.

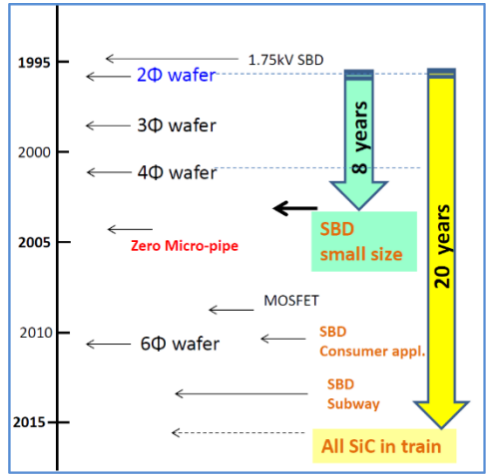

Fig.31. Timeline for development of SiC wafer node and SiC-based Schottky barrier diodes from R&D to commercial deployment [92].

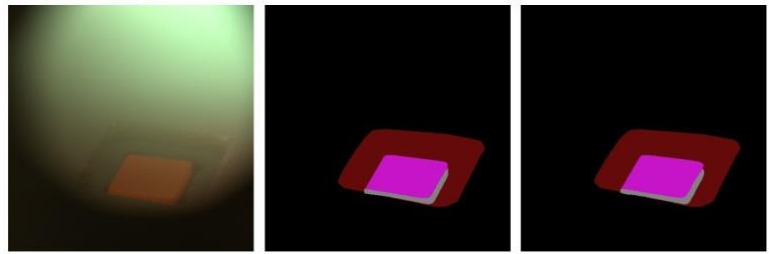

Fig.32. Visual depiction of segmentation model results on unseen test data. Left: Input Image, Middle: Actual/Labeled geometric features mask, Right: Predicted geometric features. Data presented in black, red, pink, and green represent the different independent features; background, pocket, diamond top and side.

Such an AI-guide approach was implemented into a MWCVD growth reactor for SCD wafer growth. The ultimate goal of enabling autonomous diamond growth via an AI growth control system. At such, *in situ* images of the diamond growth state and reactor telemetry were recorded. A full-frame mirrorless interchangeable lens camera (Sony Alpha 7R IV) was equipped with a macro lens (Sony SEL90M28G) for the red-green-blue (RGB) image recording. This configuration achieves pixel resolutions of ~1.6×1.6 μm$^2$. Reactor telemetry was recorded via the in-house developed diamond reactor control software. An AI algorithm pipeline was developed to model diamond growth across three interconnected thrusts of 1) Feature extraction pipeline, for extraction of geometrical features, 2) Defect detection pipeline, to extract macroscopic defect features, and 3) Frame prediction pipeline, to predict future image states 6, 8,12 and 20 hours into the future. The feature extraction pipeline concerns development of a novel DL-driven semantic segmentation approach [94]. The objective of this algorithm was to isolate and classify accurate pixel masks of geometric features like diamond, pocket holder and background, and their corresponding derivative features based on the shape and size (Fig.32). The best performing DL-based model achieved excellent feature level accuracy metrics surpassing 91.6% for pocket holder, diamond top and diamond side features of interest. Similarly, the best performing defect detection pipeline achieved accuracy metrics surpassing 92% for detecting defects forming on the growth surface of the diamond (called center defects), defects alongside the diamond edges (called edge defects) and the formation of PCD rim, whether it is on the diamond or the holder. Lastly, the geometric and defect prediction pipelines were used as inputs for the AI to generate output images. Image sequences of 5 images, each taken 10 minutes apart, and corresponding reactor telemtetry was sufficient for prediction of the growth state in the diamond reactor. Fig.33 shows such an exemplamentory growth state prediction. It can be seen that the AI-predicted (right top) image and the actual growth stage recorded by the camera 6 hours later (right bottom) have stark overlap. The AI model accurately predicted the diamond shape and size, as well as the formation of defect alongside the edge of the pocket in the pocket holder. Pixel-by-pixel prediction accuracies of >99.999% were achieved. This constitutes a never-before obtained result of spatiotemporal AI prediction of diamond shape from data obtained during a growth run, and demonstrates *in situ* growth state prediction based on a few inputs is a viable path for AI-based growth prediction, and subsequently growth control.

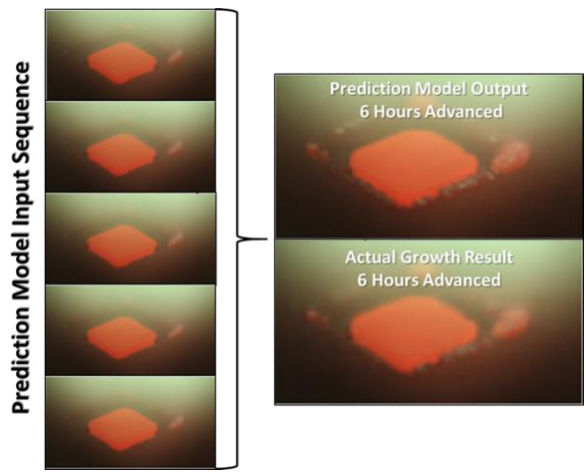

Fig.33. AI prediction of diamond shape and defects six hours ours into the future. The five images on the left serve as the input sequence. The images on the right show the comparison between the predicted (top) and actual (bottom) growth state.

The next steps will be to expand to the use of multiple reactors as to form a reactor network. Doing so will help in eliminating some of the inherent bias the current data inherently has as only a single MPACVD growth reactor is utilized. Additionally, the use of extra data will leverage the learning capabilities of the AI. Second, the move towards full integration of an automated AI-feedback governed intelligence which guides the operation of a diamond reactor will be the keystone towards unleashing the power of AI when it comes to hardware based materials development. It will leapfrog further development of diamond materials, both on the wafer level and on the application space. Additionally, it also serves as a demonstrator that AI technologies, intertwined with complex hardware, can be deployed for enhanced material development and also as much more sophisticated growth quality control tool.

**Diamond heterointegration**

Separately, the authors want to highlight another technology in which diamond has potential disruptive use, namely heterointegration of diamond with other semiconducting materials. There, diamond is co-integrated with other semiconducting materials, such as Si, SiC, GaN, and others, to build devices and solutions that leverage beneficial properties for each of the materials. For example, integration of diamond with GaN makes a lot of sense, since it benefits from the great *n*-type conductive properties in GaN, while diamond has superb thermal conduction and *p*-type conductive properties. As such diamond-GaN *p-n-*junctions are a very promising solution for heterointegrated devices. One way of achieving such a heterointegration is by using a semiconductor grafting technique to transfer an *n*-type membrane onto a prepared diamond wafer. One other specific realization of heterointegration is the creation of AlGaAs/GaAs/diamond heterojunctions for diamond-collector heterojunction bipolar transistors [95]. Generation of such semiconductor membranes, and handling thereof, are well established techniques [96]. Due to their unique properties, they can be simply transfer-printed using a polydimethylsiloxane (PDMS) stamp onto the new foreign substrate [97]. As a result, novel heterointegrated solutions will be achievable [98]. The use, and combination, of materials via membrane technology, is a potential path forward, that can bypass current limitations of heterointegration of diamond with other semiconductor materials.

**Outstanding scientific challenges**. **Plasma and reactor design and engineering**

This section reviews some outstanding issues that, in authors' opinion, are critical for advancing economic and scalable mad-made diamond production. Understanding of the interplay between the activated plasma and the substrate holder system is a key issue: it takes both the knowledge of the types and distributions of reactive species in methane/hydrogen base plasmas and the heating that activates chemistry in the gas phase and on the substrate. The interplay in question is affected by macroscopic reactor parameters such as forward microwave power, total pressure, precursor gas flow rates, sample stage cooling and overall geometry of a MPACVD reactor. Because these macroscopic parameters are simply the means to actuate plasma and reactive species and heat flow, plasma modelling becomes critical for rapid growth optimization and comparison between various reactors. Currently, state-of-the-art models try to understand SCD growth behavior by only utilizing the static reactor discharge region without the presence of the SCD sample [99, 100] or without giving details about self-consistently solving for electromagnetic (EM) field or heat transport redistributions in the presence of open or pocket holder and/or substrates [101, 102].

It must be stressed that the development of self-consistent reactor models specific to SCD growth could further reenergize the diamond supply chain technologies. Through modelling, respective plasma and substrate domains need to be optimized for high-quality growth where defect nucleation and propagation, and internal stresses due to defects and thermal gradients are inhibited, which is especially become critical when growing on the 1- to 2- inch wafer scale and beyond. Another critical aspect is that modelling must be carried out hand in hand with *in situ* reactor spectromicroscopy such that chemical species in the plasma center and periphery and thermal gradients across substrates and sample holders can be mapped and quantified. One current, basic and yet outstanding issue is benchmarking models and experiments in terms of absolute substrate temperature.

Generally, the proposed modeling approach should involve coupled steps 1) electromagnetic design of the reactor cavity, 2) plasma-microwave interaction, 3) plasma-substrate-cooling bath heat transfer, 4) chemical kinetics and reactivity of the $CH_4/H_2$ plasma, and 5) additional emerging perturbations/interactions between CVD crystal (as it grows thicker) and plasma. These effects were not fully understood. Exemplary results of such a numerical model executed in COMSOL with boundary conditions coupling , thereby completing steps 1-3, are shown in Fig.34. In this model, a pure $H_2$ plasma (pressure between 50 and 300 Torr) is ignited in a microwave cavity resonator by an external high power (1-3 kW) source operated at 2.45 GHz. $H_2$ molecules dissociate and ionize, producing electrons (at high concentrations >$10^{17}$ m$^{-3}$) absorbing the microwave power and heating up the gas to over 3,000 K. This result is highlighted in Fig.34a. Despite addition of small amounts of hydrocarbons in the feed gas (<=5:95 ratio) required for diamond growth, the main properties of the plasma discharge like thermal balance in the reactor and heat transfer to the substrate and through the sample holder/pocket (Fig.34b) are obtained by studying pure hydrogen plasma. The benefit of using this simplified situation is two-fold. First – fewer than 30 reactions need to be considered, thereby enabling a great relief in terms of computational time. Second – classical analytical plasma approaches become available, thereby providing a robust sanity-check framework to supplement the numerical results. As a final result, Fig.34c contains the calculated steady-state temperature distribution on the mosaic substrate 15 by 15 mm$^2$ surface. The model predicts correctly that the growth temperature in the substrate center is 850 °C, a typical benchmark measured across similar reactors under the same growth conditions. A somewhat negligible thermal gradient of less than 10 °C from the center to the perimeter was found. If one assumes previously considered phenomenological evidence [102] of the PCD rim formation caused by the temperature enhancement, our result may explain the PCD formation mitigation in the pocket design. In Fig.34c, one can see that at the ±8 mm mark temperature increases by a negligible amount (within computational error), while globally temperature on the outer side of the substrate is reduced. It can be naturally assumed that the temperature enhancement at the edges could be counter-balanced by the smaller heat flux from the plasma edges, thereby leading to the net zero effect and yielding uniform temperature distribution and preventing the rim formation only specific to the pocketed substrate holders.

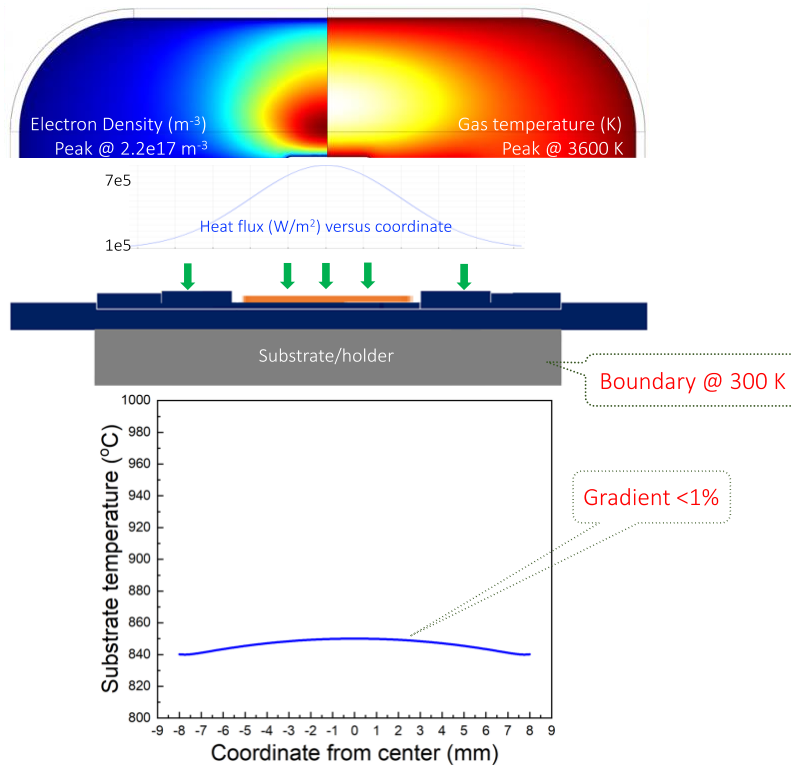

Fig.34. Developed and matured MSU high pressure (100 Torr), high power (2 kW) reactor model that self-consistently solves for a) EM-plasma power coupling and plasma heating to temperatures over 3,000 K; b) heat deposition and transfer to the substrate; and c) Resulting temperature distribution on the substrate.

Figs.15 and 24 highlighted the record growths of diamond plates of 1 inch or larger by the mosaic and heteroepitaxy techniques. On such scales, it is recorded that the growth rate become remarkably non-uniform across the plates and cracking occurs. Particularly for 2-inch growth in Fig.15, it was concluded that cracks were caused by immense internal stress as per recorded temperature difference of 100°C between the center of the samples and its edges. Since the growth rate is strongly dependent on the temperature, it decreased by about 20%. So far, growth results by the mosaic and ELO techniques or heteroepitaxy demonstrated that there is no fundamental limit on how laterally large the plate can be. The limiting factor becomes the standard reactor operating frequency of 2.45 GHz. The corresponding half-wavelength of 60 mm at this frequency implies that an alternative plasma source must be developed when approaching substrate sizes 1 inch and above. There is a great need in developing more into reactor modelling with more focus, e.g., on 915 MHz frequency standard where larger plasmas activation becomes possible. As illustrated in Fig.34, lower frequency reactors could be modelled and optimized along with pocketed sample holder configurations to achieve uniform heat flux with no thermal gradient that would yield uniform temperature that, in turn, would yield uniform growth rate and minimal internal stress while inhibiting PCD rim formation. Such reactor optimization work could therefore facilitate growing ever-larger diamond plates of the microelectronics golden standard 2 inch and larger.

Scalable production also requires enhanced growth rates without compromising quality. It has been long known that addition of only 10's ppm of nitrogen in to methane/hydrogen plasma can boost SCD growth rates ten-fold [103, 104]. Despite ubiquitous application in growing SCD, however, the causes of the growth rate enhancement remain fundamentally unclear. Obviously, $N_2$ addition to the process gas mixture affects the chemistry and composition of the microwave activated gas interacting with the substrate surface. In 2022, Ashfold and Mankelevich published a profound study on two-dimensional self-consistent modelling of N/C/H gas mixtures [105]. Using typical CVD reactor parameters, they showed that only ~0.3% of the input nitrogen molecules remain in the plasma reaction zone mostly contributing to the formation of $NH_x$ (x=0-3) and $H_xCN$ (x=0-2) species. As a result, an abysmal near surface N atom concentration was found to exist in the reactor such that the near-surface [N]/[CH3] ratio could be as low as ~$10^{-5}$. They concluded that it remained challenging to explain how such small concentrations of N-containing precursor can result in ~10-fold increases in growth rate. This calls for continuous efforts to understand fundamental plasma kinetics theoretically and through modelling that could have immediate practical growth applications.

Lastly, it is proposed that modelling using standards platforms that can be easily accessed, like COMSOL and ANSYS assisted with BOLSIG+, and shared open access reaction databases must be prioritized. This way modelling can be reproduced and verified in academic research. The importance of using codes like BOLSIG+ stems from its ability of giving a more detailed view of the electron energy distribution function (EEDF). It is generally known that, while the major fraction of electrons of roughly 1-2 eV defines properties like gas temperature, EEDF tails can drastically change plasma reactivity and reaction selectivity. Fig.35 compares methane ionization rates between Maxwellian and local EEDF approaches. More detailed analysis of Boltzmann equation for a standard MSU diamond CVD reactor demonstrated that error in ionization rate calculation, caused by the high energy EEDF tail, may be larger than a factor of 10. This will lead to errors in plasma parameters prediction in that plasma reactive species concentration calculations and hence growth rate may be incorrect. This further calls for advanced *in situ* plasma spectromicroscopy and other probe diagnostics to reveal non-Maxwellian conditions and effects.

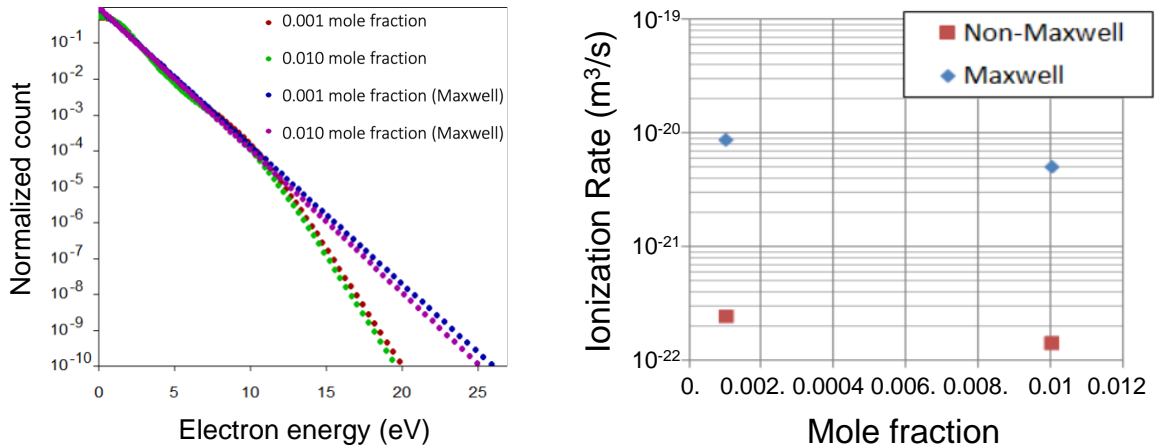

Fig.35. An example of non-Maxwellian EEDF on plasma reactivity: (left) Normalized EEDF in hydrogen plasma at two different methane mole fractions, 0.001 and 0.01 hydrogen; (right) Total ionization frequency of the $CH_4/H_2$ gas mixture at 0.001 and 0.01 mole fraction of methane.

**Conclusions**

In conclusion, an overview of recent advancements in scalable diamond production was given. In 2016, Yole Group published a graph that charted history and comparison between wafer development for the major wide bandgap semiconductors, as plotted against Si (Fig.36, top). It can be seen that progress in diamond development was curbed at no more than 1 inch. Between 2016 and 2023, true breakthroughs took place, especially in diamond heteroepitaxy. Fig.36, bottom extends the previous chart overlapping at 2015 mark. One can see that linear progress gave way to exponential growth in diamond wafer size. Projection toward 2030 predicts that this exponential trend will hold allowing for wafers as large as 6-8 inches. While this new chart is probably over-optimistic, it does highlight the fact that heteroepitaxy is expected to fuel the future progress in diamond R&D. It has to be noted that three homoepitaxy (mosaic, ELO and flipped-seed) and heteroepitaxy techniques are currently all viable options for scalable and economic production of diamond. Hence, we outline some important points when all these methods are compared in terms of scalability and economics for creating reliable supply chain for future manufacturing.

In heteroepitaxy, one big concern is the high cost of the iridium itself. Another is the lack of a process to reuse prepared wafers for multiple growth runs, like it exists in homoepitaxy, e.g., see cloning in mosaic growth. When mosaic homoepitaxy and heteroepitaxy are compared, it is clear that both techniques rely on pre-existing tile-like seeds that merge/coalesce into a single crystal under CVD growth: in mosaic method tiles are placed by hand and in heteroepitaxy tiles are generated via BEN. Because the twist, tilt and torsion set by hand will never be as perfect as can be achieved through BEN, an argument can be made that diamond grains in mosaic method will never merge perfectly at the interface but rather stay connected. As Fig.29 illustrates, in flipped seed method, the area enchantment is as good as the thickness attained in the prior growth (then flipped). As a result, the flipped side technique has not been practically implemented beyond academic feasibility studies and is considered to only serve as a fallback solution in case other techniques fall short on delivering wafers in the appropriate sizes.

It is fair to state that at this point, no technique can achieve size, quality and quantity at the same time. Wafer size in heteroepitaxy is the largest and so is the defect density, $10^7$-$10^9$ cm$^{-3}$ [106], much higher compared to those in homoepitaxy. Such high defect densities compromise application of heteroepitaxially grown SCD in electronics. While area enlargement in ELO is modest, defect densities in the low $10^3$ cm$^{-3}$ are routine [107]. When ELO is used together with additional substrate engineering techniques [108, 109], it is possible to reduce the defect density down to $10^2$ cm$^{-3}$ [110], a value otherwise only achieved by the highest quality HPHT crystals. In Fig.28, a striking example of dislocation reduction in the flipped seed method is highlighted. It can be hypothesized that, moving forward, new combinations between the existing four methods could be developed to pave ways for new methodologies attaining large size, low defect density and low-cost diamond wafers.

From historic perspective, the progress in diamond synthesis consisted from two stages: 1) development of physics and chemistry of methane/hydrogen plasma activation and kinetics that led to engineering development of the workhorse reactor designs and 2) engineering development of novel substrate and substrate holder techniques, like ion-assisted lift-off, BEN, mosaic assembly and pocket holder. In this paper, it was argued that in order to enable future progress and diamond manufacturing, self-consistent synergy should be applied between plasma modelling and theory and *in situ* diagnostics, and substrate/holder optimization. This way best conditions could be found and implemented where chemical species and heat flux distributions are of proper magnitude and uniformity to enable 2+ inch wafers growths at high rates and minimal thermal stress mitigating cracking and minimizing dislocation densities. When models, *in situ* spectromicroscopy data and other reactor observables are combined with AI and machine learning, it is expected to further bolster advancement in diamond manufacturing for future microelectronics that includes both all-diamond and heterointegrated devices.

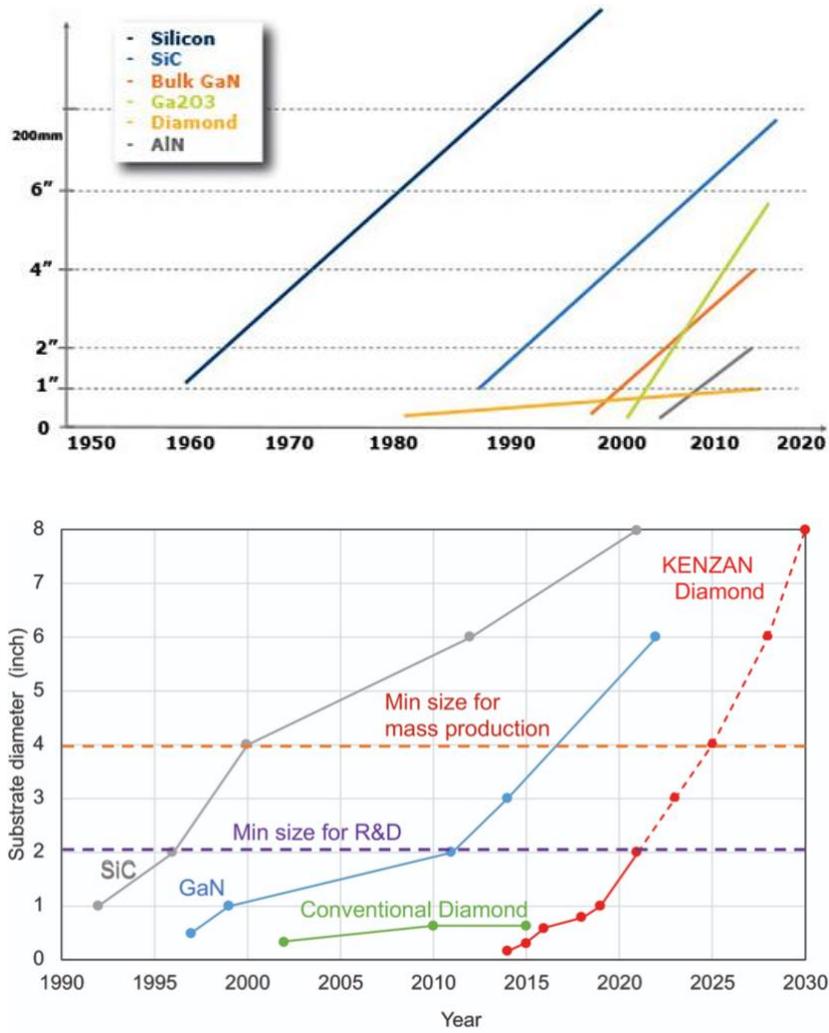

Fig.36. Historic charts showing the progress in diamond substrate size.

**Acknowledgment**

This work was supported by internal funds provided by College of Engineering, Michigan State University and Fraunhofer USA, Inc., NSF award 2036737 including a supplemental INTERN award, and DOE Basic Energy Science award DE-SC0020671.